# Design of an Improved Microstrip Antenna Operating at a Frequency Band of 28 GHz

S. O. Zakariyya[1*], V. O. Omole[1], B. O. Sadiq[2,3], R. A. Alao[1], J. A. Adesina[4], E. Obi[5]

[1]Electrical and Electronics Engineering Department, University of Ilorin, Ilorin, Nigeria.
[2]Electrical and Computer Engineering, Kampala International University, Uganda
[3]Computer Engineering Department, Ahmadu Bello University, Zaria, Nigeria
[4]Computer Engineering Department, University of Ilorin, Ilorin, Nigeria.
[5]Electronics and Telecommunication Engineering Department, Ahmadu Bello University, Zaria, Nigeria
*Email: zakariyya.os@unilorin.edu.ng

*ABSTRACT*
The design of an improved microstrip antenna operating in the 28 GHz frequency spectrum is the main goal of this work. The design used a Roger RT 5880 LZ substrate with a thickness and permittivity of 0.762mm and 1.96, respectively. The antenna was simulated in CST Microwave Studio. As the antenna feed, a quarter-wave transformer was used to provide an impedance match of 50 ohms. To improve the antenna's performance, a U-shaped element was added to the ground plane. The antenna resonated at 28 GHz frequency, according to simulation data, with a return loss of -21.4 dB, VSWR of 1.18, bandwidth of 2.026 GHz, and gain of 8.19 dB. The proposed antenna exhibits a performance improvement in terms of gain and bandwidth due to the addition of U-shaped element when benchmarked with existing designs in the literature work.

## 1. INTRODUCTION

The evolution of mobile and cellular networks has experienced remarkable progress over recent decades, evolving through various generations that each introduced enhanced features surpassing their predecessors. Demand for bandwidth and mobile broadband has surged exponentially in the last decade, resulting in a significant increase in data traffic that largely outpaced the capabilities of prior generations, most notably 4G. This burgeoning demand, particularly for data-intensive applications, necessitated a shift towards higher frequencies beyond the traditionally used sub-6 GHz range before the advent of 5G. These earlier frequencies, though sufficient for past needs, began to experience congestion issues and network capacity limitations, underscoring the imperative for higher bandwidth and data rates to support the emerging landscape of digital consumption and services (Zakariyya et al., 2019).

5G technology has been designed to address these challenges by not only utilizing the existing sub-6 GHz frequency bands but also expanding into the millimeter-wave spectrum, which ranges from 24 GHz to 100 GHz. This shift to higher frequencies is a cornerstone of 5G's ability to provide higher data rates, reduced latency, and increased capacity and the ability to connect numerous devices simultaneously. The millimeter waves frequencies, despite their limited propagation range and higher susceptibility to both attenuation and environmental obstructions, unlocks the potential for significantly expanded bandwidth and capacity. Among the existing millimeter waves spectrum, 28-GHz is notably preferred for 5G applications due to its lower atmospheric absorptions and attenuations rates (Larsson et al, 2014; Hu et al, 2018).

The design of antennas, especially Microstrip Patch Antennas (MPAs), is crucial in overcoming the limitations of the millimeter wave band and maximizing the use of this wider frequency range. MPAs offer significant advantages due to their compact size, design flexibility, and seamless integration with diverse devices and network setups (Zakariyya et al., 2015; Zakariyya et al., 2016; Salami et al., 2018). However, traditional patch antennas often suffer from low gain and limited bandwidth. To address these shortcomings, this study introduces an improved patch antenna tailored for 5G communication.

Rahman et al. (2016) presented an antenna operating at 28 GHz with a bandwidth of 2.66 GHz. However, this design lacked compactness and had a low gain. Przesmycki et al. (2020) designed a rectangular patch antenna operating at 28 GHz, utilizing RT/duroid substrate. This antenna achieved notable metrics. However, the gain did not meet the requirements for 5G applications. Awan et al. (2021) discussed a microstrip antenna with a defective ground structure (DGS) at 28 GHz. While this design exhibited good bandwidth and return loss, its gain may still be insufficient to overcome the path and absorption losses in the millimeter-wave spectrum.

Kamal et al. (2021) introduced a single-band hook-shaped antenna at 28 GHz. This antenna demonstrated a good return and wide bandwidth. However, its size was not compact, and its gain was relatively low. Additionally, Raheel et al. (2021) proposed a microstrip patch antenna operating at 28/38 GHz. This design achieved a gain of 7.1 dBi and an impedance bandwidth of 1 GHz (27.6 GHz - 28.6 GHz). Hussain et al. (2022) attempted a circular microstrip patch antenna with two rectangular slots, showcasing a good impedance bandwidth. However, this antenna was relatively large and had a modest gain. Furthermore, Farahat & Hussein (2022) introduced a patch antenna operating in the 28/38 GHz bands, achieving a gain of 6.6 dB and a bandwidth of 1.23 GHz in the 28 GHz range. Gaid et al, 2024 proposed a microstrip patch antenna operating at 28/38 GHz. This design achieved a gain of 8.1 dB and an impedance bandwidth of 1.43 GHz.

These discussed antennas exhibit various strengths and weaknesses, such as wide bandwidths, high gains, or compact sizes. However, their performance can be improved by adding a U shape element on the ground plane. Hence, this research aims to strike a balance between antenna size, impedance bandwidth, and gain by designing a small-size antenna with high gain, optimal impedance bandwidth, and coverage of the 28 GHz band.

## 2. ANTENNA DESIGN

The initial step in creating a patch antenna involves selecting the substrate, as its properties, like thickness and dielectric constant, significantly impact the antenna's characteristics. In this particular design, a Roger RT 5880LZ substrate was chosen with a thickness of 0.762mm and a relative dielectric constant of 1.96. Copper was used for both the patch conductor and ground plane. These materials were chosen due to their high conductivity. The dimensions of the patch element are calculated using transmission line model accordingly (Zakariyya et al., 2016; Balanis, 2005):

$$w_{pc} = \frac{v}{2f_r}\sqrt{\frac{2}{\varepsilon_{r\_l}+1}} \quad (1)$$

The patch width is $w_{pc}$, $\varepsilon_{r\_l}$ is the relative dielectric constant and $f_r$ represents the frequency at which the antenna resonates and $v$ is the speed of light.

The effective dielectric constant is expressed as where $h_{\_s}$ is substrate height:

$$\varepsilon_{r\_eff} = \frac{\varepsilon_{r\_l}+1}{2} + \frac{\varepsilon_{r\_l}-1}{2}\left(1+\frac{12h_{\_s}}{w_{pc}}\right)^{\frac{-1}{2}} \quad (2)$$

As a result of fringing, the length of the patch is extended by a distance equal to:

$$\Delta L_{\_p} = 0.412 h_{\_s} \frac{(\varepsilon_{r\_eff}+0.3)(w_{pc}/h_{\_s}+0.264)}{(\varepsilon_{r\_eff}-0.258)(w_{pc}/h_{\_s}+0.8)} \quad (3)$$

The length is determined by:

$$L_{\_p} = L_{\_eff} - 2\Delta L_{\_p} \quad (4)$$
$$L_{\_eff} = \frac{v}{2f_r\sqrt{\varepsilon_{r\_eff}}} \quad (5)$$

The edge impedance $Z_{in} = 90 \frac{\varepsilon_{r\_l}^2}{\varepsilon_{r\_l}-1}\left(\frac{L_{\_p}}{w_{pc}}\right)^2 \quad (6)$

Where $L_{\_p}$ is the patch length

The line impedance of the quarter wave is given by:

$$Z_T = \sqrt{Z_0 Z_{in}} \quad (7)$$

The length ($L_{\_p}$) and width ($w_{pc}$) of the radiating patch for the design was calculated to be 3.2 mm and 4.4 mm using the mathematical equations described in (1-4). The load impedance at the edge of the patch is calculated to be approximately 190.5 ohms using Eq. (6). A simulation model for the single element antenna implemented in CST-MWS is shown in Figure 1. For enhancement of the antenna performance, a U shape element is created on the ground plane to serve as parasitic element.

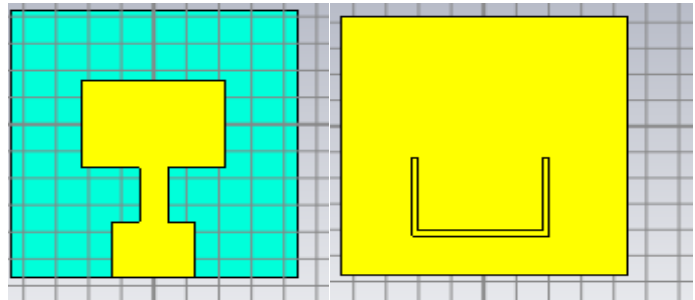

Figure 1: Front and back view of the designed antenna

## 3. RESULTS AND DISCUSSION

### 3.1. Return Loss

Figure 2 shows the return loss of the proposed MPA. For mobile communication systems, a standard value of -10 dB is acceptable as the baseline for good performance. The proposed MPA has a return loss of -21.4 dB, resonating at the desired resonant frequency of 28 GHz and covering a frequency band of 27.185 GHz – 29.211 GHz with a bandwidth of 2.026 GHz.

### 3.2. Voltage Standing Wave Ratio

Figure 3 shows the Voltage Standing Wave Ratio (VSWR) of the proposed MPA design. For mobile communication systems, the value of VSWR should be small (not more than 2.5 and close to 1.0) in order to ensure proper impedance matching and little amount of reflected power. The proposed MPA design achieved a VSWR of 1.18 at a resonant frequency of 28 GHz which indicates good matching between the feeding line and the radiating patch element.

### 3.3. Radiation Pattern

Figure 4a and Figure 4b shows the 2D and 3D radiation pattern plot of the proposed MPA design. As observed from the 3D plot, the proposed patch antenna design has a gain of 8.19 dBi which is considered acceptable in this scenario of compact antenna design.

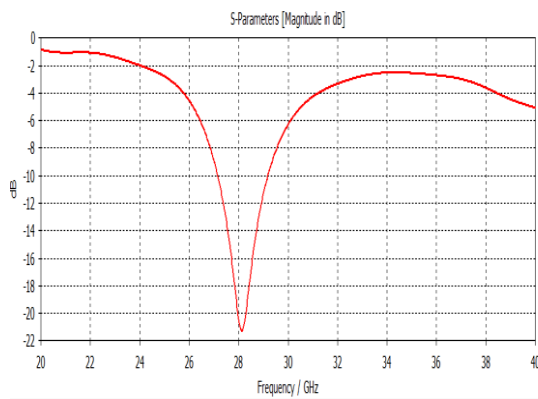 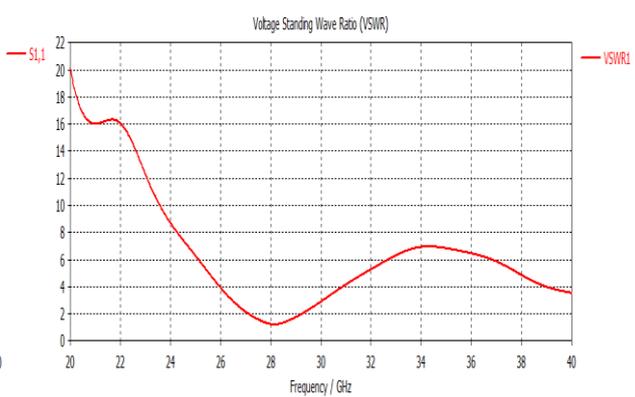

Figure 2: Return loss plot             Figure 3: voltage standing wave ratio

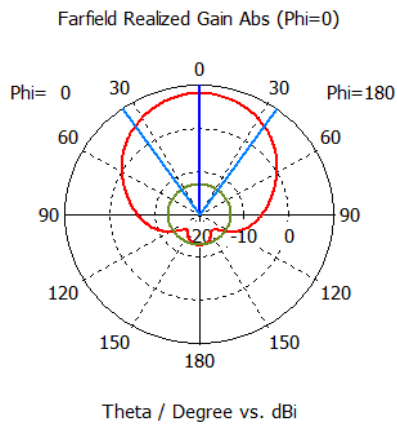 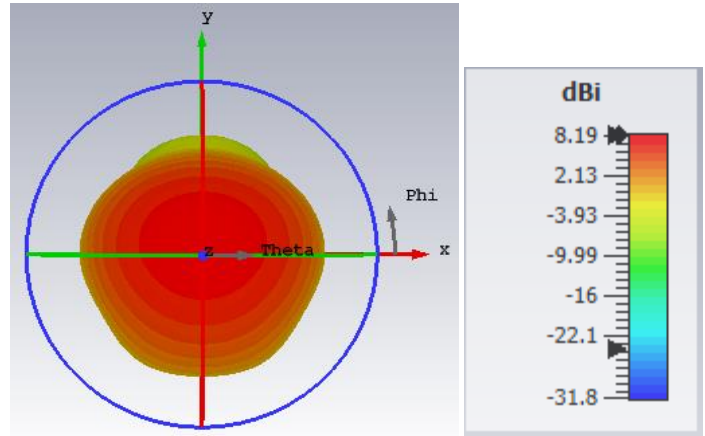

Figure 4a: 2D Radiation pattern plot      Figure 4b: 3D Radiation pattern plot

Table 1 is a summary of the comparison between the proposed design and other reported 5G antennas. From the table, the proposed 5G antenna antenna is found to perform better than the majority of the counterparts in terms of the bandwidth and gain.

Table 1: Comparison between the Proposed Design and Other Reported Antennas

| Reference Antennas | S11(dB) | Bandwidth (GHz) | Gain (dB) |
|---|---|---|---|
| Proposed design | -21.4 | 2.02 | 8.19 |
| Gaid et al., 2024 | -45 | 1.43 | 8.1 |
| Farahat & Hussein, 2022 | -34.5 | 1.23 | 6.6 |
| Raheel et al., 2021 | -25 | 1 | 7.1 |

## 4. CONCLUSION

A microstrip patch antenna has been designed for 5G applications in mobile communication. The antenna was designed to resonate at 28 GHz. A U-shape structure was added to the ground plane with the aim of enhancing the antenna's performance. The proposed MPA simulation results show that the return loss, bandwidth, gain and VSWR are -21.4 dB, 2.02 GHz, 8.19 dB, and 1.18, respectively. The proposed antenna offers highly competitive performance when compared to other designs. The suggested structure is a good candidate for 28 GHz band applications due to its simplicity, good radiation properties, and compactness